# Standing-Wave Optical Trap Based on Retro-Reflection Photonic Nanojet

Yu.E. Geints[1], I.V. Minin[2], O.V. Minin[2]

**Abstract**: A concept of an innovative optical trap based on the retro-reflected standing-wave photon nanojet (SWOT) is presented. An open resonance cavity is formed between two coaxial microparticles of different geometries (sphere, cylinder, ring, truncated cone) with one particle docked to a plain mirror. Numerical simulations have shown the achievement of a record-high optical field intensity in the SWOT workspace, almost seven times higher than that of a conventional photonic nanojet trap due to a triple-focused optical beam, which contributes to improved optical capture. The proposed design of the optical trap allows for multi-position particle confinement in the trap area. The advantages of the proposed solution are the simple technical implementation and the possibility of integration with microfluidic technologies for optical manipulation of nanoobjects (Chip-on-flex optical sorting and ordering of nanoobjects, particle beaming).

The nature of light and the understanding of the essence of optical phenomena have interested researchers throughout the entire history of human evolution. A key element in comprehending the properties of light is explaining how light interacts with matter. In 1899, Lebedev, and independently Nichols and Hull in 1901, experimentally demonstrated that light exerts pressure on small suspended plates [1]. This discovery later led to the development of many fundamental areas in physics. Research on laser light pressure has been recognized with several Nobel Prizes in Physics [2]. In applied fields, the most widespread studies include, for example, the design of solar sails [3] and optical tweezers [4], which continue to be improved [1, 5].

Among the wide variety of principles for constructing optical tweezers [6], an optical trap (OT) based on standing waves is particularly attractive. Fundamentally, such a trap enables control over a chain of nanoparticles positioned at the extrema of the interference field of a standing wave [7]. Over the past two decades, there has been an increase in scientific interest in mesoscale dielectric particles with effective dimensions comparable to the wavelength of the illuminating radiation. This is due to their unique property of modifying the spatial structure of the incident optical wave and creating a concentrated subwavelength light flux with enhanced intensity in the near-field scattering region within the shadow part of the particle. This specific region, typically elongated along the direction of incident radiation, is referred to in the literature as a «photonic (nano)jet» (PNJ) [8]. These unique structural features of photonic jets are precisely what enable broad opportunities for their practical application in various fields of science and optical technologies [9–11]. A formal relationship between the optical force acting on such a spherical particle and the photonic jet it generates was demonstrated in [12].

Optical traps based on standing waves are of particular interest for practical tasks, as the spatial localization of the optical field is formed not only in the transverse but also in the longitudinal directions, amount packed to half the wavelength in the medium [13]. Two types of such PNJ-traps are typically considered: a trap based on counter-propagating PNJs with two radiation sources [14], and a trap with a single radiation source and a mirror [15].

An optical trap based on two interfering photonic jets, created by two particles illuminated from opposite directions by two different radiation sources, was examined in [14, 16]. Meanwhile, an optical trap based on a standing wave with a dielectric mesoscale microparticle positioned near a reflecting mirror and irradiated by a single optical source (retro-OT) was proposed in [15]. The refinement of the optical trap based on the photonic jet with mirror reflection and a single optical source enabled the proposal of an optical trap with improved parameters [17]. The dynamic characteristics of an optical trap based on a photonic jet modulated by a standing wave are considered in [18]. Nevertheless, this relatively new and promising type of OT has not yet been sufficiently investigated, and its potential capabilities have not been covered in the literature.

On the other hand, another type of mesoscale photonic devices for optical trapping and manipulation of nanoparticles and atoms is based on microresonators [19, 20]. It appears beneficial to combine these two approaches for implementing optomechanical manipulation of objects [21], which, to our knowledge, has not yet been addressed in the photonic community.

In present work, building on the results of study [17], we thoroughly examine the structure of the optical field of a retro-reflected PNJ formed in an open resonant system consisting of a pair of dielectric (non-absorbing) microparticles and a flat mirror. We propose a novel multi-position retro-OT on a standing wave (SWOT) with significantly improved optical trapping parameters. It should be noted that the proposed OT introduces an additional degree of freedom, which consists in the ability to change the shape of focusing particles (not limited to spheres or cylinders). This will enable control over the characteristics of the standing wave field, and consequently, over the trapping parameters of the optical trap itself.

The physical concept of the proposed OT is illustrated in Fig. 1a. SWOT is formed by a pair of mesoscale dielectric particles (their shape can vary) spaced apart from each other to create a region of optical trapping between them in the form of an open resonant cavity with a leaky mode. In turn, to generate such a mode, i.e., a standing wave with equidistant antinodes and nodes of optical intensity, the upper microparticle (in Fig. 1a) is "docked" from below to a flat mirror. This mirror is fabricated, for example, from a metal foil (Au-film) on a polymeric spacer. The entire structure is mounted on a transparent substrate and illuminated from below by an optical wave (magenta arrow) directed towards the mirror.

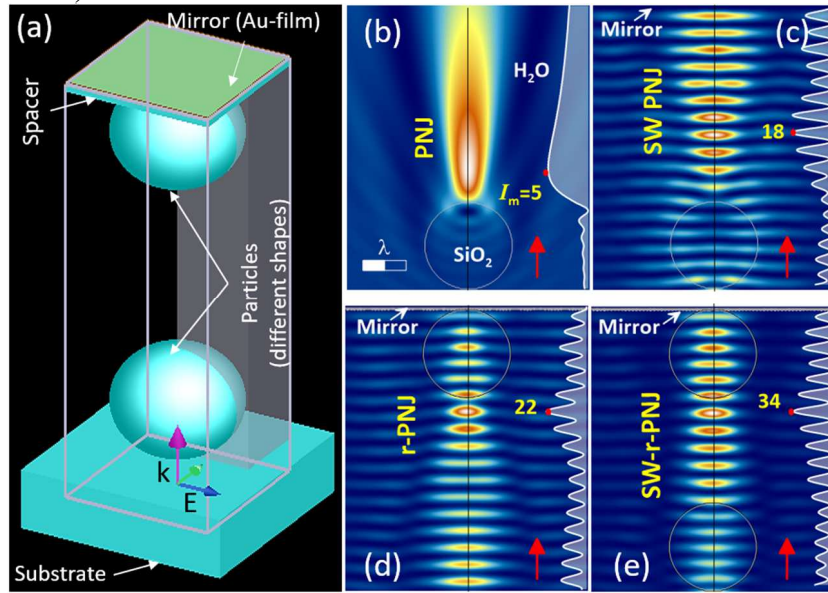

Fig. 1. (a) A perspective view of the multi-position SWOT. (b-e) Distribution of the square of optical field amplitude (intensity) $|\mathbf{E}|^2$ in the cross section of (b) regular PNJ, (c) standing wave PNJ (SW-PNJ) and (d) retro-reflected PNJ (SW-PNJ), as well as (e) in a retro-reflected PNJ supported by a standing wave (SW-r-PNJ). The maximum intensity ($I_m$) is indicated in yellow numbers.

To understand the operating principle of the proposed SWOT, let us examine Figs. 1b–e which present 2D distributions of the relative (normalized) optical intensity $I = |\mathbf{E}|^2$, where $\mathbf{E}$ is the electric field amplitude, in the polarization plane of the incident optical wave ($x$–$z$). These distributions sequentially illustrate the evolution of the OT based on a photonic jet, starting from a configuration with a regular PNJ produced by a single glass microsphere (Fig. 1b) and ending with a SWOT combination of two spheres and a mirror (Fig. 1e). On all distributions, the axial intensity profile is shown on the right, and the maximum intensity value achieved in this type of OT (under the parameters specified below) is indicated by yellow numbers.

Numerical field simulations are performed using the finite-difference time-domain (FDTD) method implemented in the Ansys Lumerical FDTD Electromagnetic Solver environment. A numerical solution of the vector Helmholtz equation for the electromagnetic field components in the steady state was carried out [22]: $\nabla \times \nabla \times \mathbf{E}(\mathbf{r}) - k^2 \varepsilon(\mathbf{r}) \mathbf{E}(\mathbf{r}) = 0$, where $\varepsilon(\mathbf{r})$ is the dielectric permittivity of the medium (dependent on coordinates), $k = 2\pi/\lambda$ is the wavenumber, and $\lambda$ is the optical wavelength. The optical radiation is assumed to be monochromatic, with a telecommunication wavelength $\lambda = 1550$ nm. The photonic structures under study were placed in water with a refractive index $n = 1.318$ [23]. It is assumed that there is no optical absorption in any materials or media, except for the mirror, which is modelled as a gold foil [24].

The substrate and spacer material are chosen to be fused silica (SiO$_2$) with $n = 1.44$, while the microparticles are fabricated from optical glasses with a refractive index ($n$) varying in the range from 1.44 to 1.7. The entire computational domain is bounded by perfectly matched layers (PMLs) to prevent parasitic reflection of optical waves from the domain boundaries. An adaptive discretization mesh is used, with a minimum spatial step of 2.5 nm and a temporal step of 0.1 fs. The particle parameters are specified below in the description of Fig. 2.

Returning to Fig. 1, we can see that in the most common case of PNJ formation in the shadow region of the near-field of a single glass microsphere with $n = 1.5$ (Fig. 1b), the photonic jet is characterized by a smooth hill-shaped intensity profile with a slight intensity increase, $I_m = 5$. Here, the longitudinal scale of intensity variation is rather large, amounting to several radiation wavelengths.

If a flat mirror is placed across the direction of photonic jet propagation, as shown in Fig. 1c, one obtains an open resonator with asymmetric wave reflection that supports leaky eigenmodes. In this case, a standing wave is formed in the resonant cavity through the constructive interference of the direct and counter-reflected PNJs (SW PNJ). Due to the presence of the standing wave, the peak intensity in this configuration is nearly four times higher, reaching $I_m = 18$. Additionally, longitudinal discontinuity of the PNJ emerges, with a period that is a multiple of the optical half-wavelength in the medium, i.e., $\Delta z = \lambda/2n \sim 0.6$ μm.

Further enhancement of the optical field in the PNJ region can be achieved using the retro-reflection design (r-PNJ) [15, 17], where the microparticle generating PNJ is placed directly against the mirror, as shown in Fig. 1d. In this case, unlike in Fig. 1c, the dielectric particle focuses the optical radiation twice — before and after reflection from the mirror — before it interacts with the incident field. As a result of this double focusing, a slight increase in intensity within the standing wave is observed, reaching the value $I_m = 22$. Note that in this case, the resonator properties of the photonic structure are lost.

Finally, the combination of the resonant SW PNJ and retro-reflector r-PNJ schemes results in a retro-PNJ configuration with a standing wave (SW-r-PNJ), as shown in Fig. 1e. The optical radiation incident on the system of two particles and a mirror is first focused by the initial particle, forming an extended focusing region (PNJ). This focused beam is then refocused by the second particle and reflected from the flat mirror in the opposite direction, returning to the same (second) particle. This already triple-focused optical beam interacts with the primary PNJ (rather than with the incident field), and through constructive interference, generates a periodic sequence of intensity maxima within the resonant cavity between the particles. Worthwhile noting, such interference of optically beams pre-concentrated into photonic jets leads to a significant increase in intensity in the region between the particles, now reaching $I_m = 34$ that is nearly seven times higher than maximal intensity enhancement in a conventional PNJ. We also note that this OT configuration uses only a single optical beam, rather than two beams as in known trapping schemes with counter-propagating colliding PNJs [13, 14].

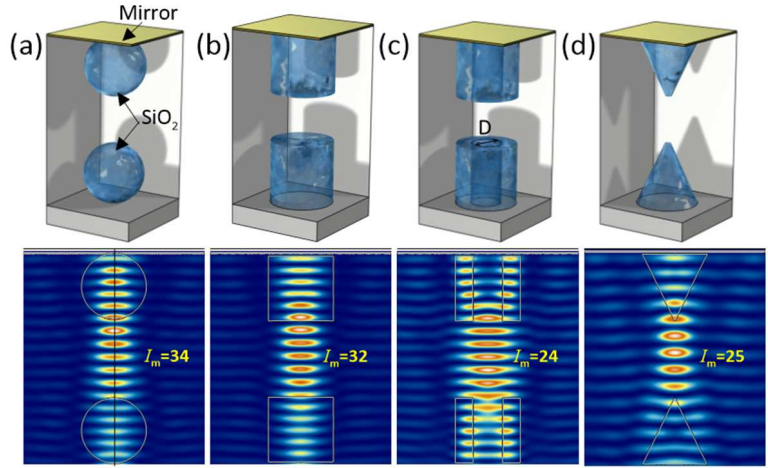

Fig. 2. SWOT with various shapes of working microparticles, (a) sphere, (b) circular cylinder, (c) ring, (d) circular truncated cone. The optical intensity distribution is shown below.

The open nature of the discussed OT allows using of microparticles of various shapes as elements focusing the optical wave. In particular, we examined four types of SWOTs based on pairs of glass spheres, cylinders, rings, and truncated cones, as shown in Figs. 2a–d. The figure also presents the structure of optical intensity in the optical trapping region of the trap. All particles have identical linear dimensions (base diameter, height) equal to $2\lambda$, and a refractive index $n = 1.7$. The inner hole diameter $D$ of the circular ring (Fig. 2c) is $0.7\lambda$. It can be seen that, in principle, all OT types are characterized by a sufficiently high intensity enhancement in the resonant cavity with a standing wave. At the same time, particles with more complex geometric shapes, namely, the ring and cone, result in lower field concentration due to the reduced lensing effect when the optical wave reflects off the mirror.

It is important to evaluate the optical trapping potential of the proposed OT based on the retro-PNJ standing wave effect. As is known, the mechanical force acting on a probe particle from the optical field can be calculated based on the Maxwell stress tensor and the normal component of the Poynting vector $\mathbf{P}$ [22]. Formally, the optical force is the resultant of two forces: $\mathbf{F} = \mathbf{F}_s + \mathbf{F}_\nabla$, where $\mathbf{F}_s$ is the scattering force, arising from light pressure on the particle surface, and $\mathbf{F}_\nabla$ represents the gradient optical force, associated with the Lorentz force acting on a charge in an electric field. The scattering force acts in the direction of optical wave propagation, while the gradient force is directed opposite to the surface-averaged gradient of the electric field vector.

For Rayleigh particles with characteristic dimensions much smaller than the radiation wavelength, when the dipole (electrostatic) approximation can be used, analytical expressions for these forces exist in the following form [4]: $\mathbf{F}_s = \mathbf{e}_k \alpha^2 k^4 I / 6\pi c$, and $\mathbf{F}_\nabla = -\alpha (\nabla I)/cn\varepsilon_0$. Here, $\alpha$ is the polarizability of the test particle, $I = (cn\varepsilon_0/2)|\mathbf{E}|^2$ is the optical intensity, $\mathbf{e}_k$ is the unit vector in the direction of radiation propagation, $\nabla$ denotes the vector operation of spatial field gradient, and $\varepsilon_0$ is the vacuum permittivity.

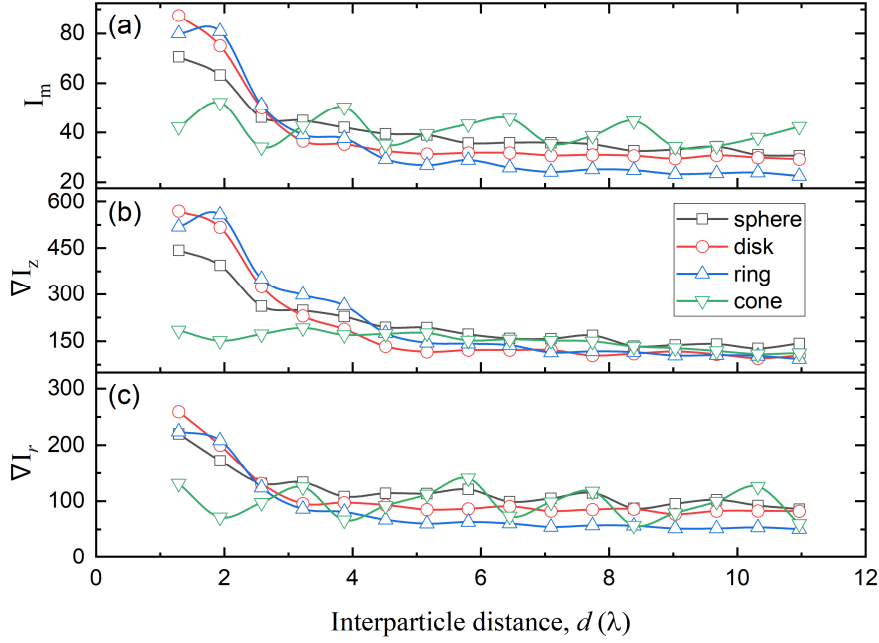

Fig. 3. Optical SWOT parameters. (a) Maximum intensity $I_m$, (b) longitudinal $\nabla I_z$ and (c) transverse intensity gradient $\nabla I_r$ depending on interparticle distance $d$ and particle type.

One of the key features of optical traps utilizing the standing-wave effect (colliding PNJs, retro-reflection PNJ, and SWOT) is the significantly reduced optical force arising from light scattering on the particle. This is because the presence of counter-propagating optical fluxes that diminishes the longitudinal component of the Poynting vector $P_z$ [4]. Clearly, under these conditions, the role of the gradient optical force $\mathbf{F}_\nabla$ in the optical trapping capability increases. This is further enhanced by the additional rise in intensity at the antinodes of the optical field in standing-wave-type traps, compared to a classical PNJ trap (Fig. 1b). To characterize the trapping potential of the proposed SWOT, we calculate the maximum value of the optical field intensity gradient in the longitudinal $\nabla I_z = \lambda(\partial I/\partial z)$ and transverse $\nabla I_r = \lambda(\partial I/\partial x + \partial I/\partial y)/2$ directions (at the intensity maximum), depending on the particle shape and the interparticle gap $d$ which bounds the OT working region. The simulation results are presented in Fig. 3.

By analyzing this figure, two characteristic regions can be distinguished, which show different trends of field intensity parameters. There are, (a) the "close-contact region" of microparticles in the OT, where the gap between particles does not exceed ~ 3λ, and (b) the "far-contact region", when $d > 3\lambda$. In the close-contact region, only a few antinodes are formed in the standing wave typically yielding just one or two zones of stable optical trapping. Here, the intensity gradients reach their maximum values, and thus the trap stiffness is maximal. It can be seen that among all particle types, those based on a cylindrical column (cylinder and ring) exhibit a clear advantage in terms of field gradient magnitude. The lowest optical trapping forces are supposed to occur in SWOTs composed of a pair of conical particles, which is evidently due to the small area of the particle output (focusing) surface.

As the particles are moved apart, all significant field parameters within the trapping region gradually decrease. However, the number of intensity maxima in the standing wave within the trapping region increases proportionally, enabling multi-position optical confinement of several test particles simultaneously. In the far-contact region, none of the microparticle shapes exhibits a clear advantage in longitudinal trapping showing close values in $\nabla I_z$. Interestingly, SWOT based on two spheres demonstrates a consistently high transverse intensity gradient $\nabla I_r$, while the trap formed by a pair of rings exhibits the weakest transverse confinement, due to the stronger angular divergence of the resulting PNJs.

In conclusion, we have presented a new concept of an optical trap based on the retro-reflected standing-wave photonic nanojet effect (SW-r-PNJ), which arises in the gap between two coaxial micron-scale dielectric particles (sphere, cylinder, ring, cone, etc.), one of which is placed near a mirror-reflecting surface. A detailed analysis of the near-field optical structure of such a SW-r-PNJ reveals that a system of two dielectric microparticles and a flat mirror forms a distinctive open resonator supporting leaky eigenmodes in the form of a standing wave. These modes are characterized by a higher concentration of optical radiation in the interference region of the direct and reflected photonic jets. This endows the retro-PNJ standing-wave OT with several unique properties distinct from previously known PNJ-based optical traps. In particular, the proposed OT type should be characterized by significantly higher optical trapping forces and extended multi-position capability as the working zone length (inter-particle gap size) increases.

We believe that, due to its relatively simple technical implementation, the proposed OT can find practical application as a Chip-on-flex microfluidic device [25] for optical sorting and beaming (line-up) of target atoms or nanoscale objects [26]. Moreover, SWOT could enable combining multi-position optical manipulation with a microfluidic channel, for instance, to localize and hold various nanoscale objects prior to their removal from the target region [27, 28]. We anticipate that the proposed concept will lead to further integration of optomechanical trapping with microfluidic technologies, and will expand the degrees of freedom in optical manipulation through precise tuning of the localized field optical parameters.